\documentstyle[]{mn}
\newif\ifAMStwofonts

\ifoldfss
  \newcommand{\rmn}[1] {{\rm #1}}

  \ifCUPmtlplainloaded \else
    \NewTextAlphabet{textbfit} {cmbxti10} {}
    \NewTextAlphabet{textbfss} {cmssbx10} {}
    \NewMathAlphabet{mathbfit} {cmbxti10} {} 
    \NewMathAlphabet{mathbfss} {cmssbx10} {} 
  \fi
  \ifAMStwofonts
    \ifCUPmtlplainloaded \else
      \NewSymbolFont{upmath} {eurm10}
      \NewSymbolFont{AMSa} {msam10}
      \NewMathSymbol{\upi}     {0}{upmath}{19}
      \NewMathSymbol{\umu}     {0}{upmath}{16}
      \NewMathSymbol{\upartial}{0}{upmath}{40}
      \NewMathSymbol{\leqslant}{3}{AMSa}{36}
      \NewMathSymbol{\geqslant}{3}{AMSa}{3E}

    \fi
  \fi
\fi 

\ifnfssone
  \newmathalphabet{\mathit}
  \addtoversion{normal}{\mathit}{cmr}{m}{it}
  \addtoversion{bold}{\mathit}{cmr}{bx}{it}
  \newcommand{\rmn}[1] {\mathrm{#1}}

  \newmathalphabet{\mathbfit} 
  \addtoversion{normal}{\mathbfit}{cmr}{bx}{it}
  \addtoversion{bold}{\mathbfit}{cmr}{bx}{it}
  \newmathalphabet{\mathbfss} 
  \addtoversion{normal}{\mathbfss}{cmss}{bx}{n}
  \addtoversion{bold}{\mathbfss}{cmss}{bx}{n}
  \ifAMStwofonts
    \ifCUPmtlplainloaded \else
      \UseAMStwoboldmath
      \makeatletter
      \new@mathgroup\upmath@group
      \define@mathgroup\mv@normal\upmath@group{eur}{m}{n}
      \define@mathgroup\mv@bold\upmath@group{eur}{b}{n}
      \edef\UPM{\hexnumber\upmath@group}
      \new@mathgroup\amsa@group
      \define@mathgroup\mv@normal\amsa@group{msa}{m}{n}
      \define@mathgroup\mv@bold\amsa@group{msa}{m}{n}
      \edef\AMSa{\hexnumber\amsa@group}
      \makeatother
      \mathchardef\upi="0\UPM19
      \mathchardef\umu="0\UPM16
      \mathchardef\upartial="0\UPM40
      \mathchardef\leqslant="3\AMSa36
      \mathchardef\geqslant="3\AMSa3E
    \fi
  \fi
\fi 

\ifnfsstwo
  \newcommand{\rmn}[1] {\mathrm{#1}}

  \DeclareMathAlphabet{\mathbfit}{OT1}{cmr}{bx}{it}
  \SetMathAlphabet\mathbfit{bold}{OT1}{cmr}{bx}{it}
  \DeclareMathAlphabet{\mathbfss}{OT1}{cmss}{bx}{n}
  \SetMathAlphabet\mathbfss{bold}{OT1}{cmss}{bx}{n}
  \ifAMStwofonts
    \ifCUPmtlplainloaded \else
      \DeclareSymbolFont{UPM}{U}{eur}{m}{n}
      \SetSymbolFont{UPM}{bold}{U}{eur}{b}{n}
      \DeclareSymbolFont{AMSa}{U}{msa}{m}{n}
      \DeclareMathSymbol{\upi}{0}{UPM}{"19}
      \DeclareMathSymbol{\umu}{0}{UPM}{"16}
      \DeclareMathSymbol{\upartial}{0}{UPM}{"40}
      \DeclareMathSymbol{\leqslant}{3}{AMSa}{"36}
      \DeclareMathSymbol{\geqslant}{3}{AMSa}{"3E}
    \fi
  \fi
\fi 

\ifCUPmtlplainloaded \else
  \ifAMStwofonts \else 
    \def\upi{\pi}
    \def\umu{\mu}
    \def\upartial{\partial}
  \fi
\fi

\newcommand{\COBE}
        {{\em COBE}}

\newcommand{\cdm}
        {CDM}

\newcommand{\cmb}
        {CMB}

\newcommand{\EdS}
        {EdS}

\newcommand{\JVAS}
        {JVAS}

\newcommand{\TdF}
        {2dF}

\newcommand{\LBQS}
        {LBQS}

\newcommand{\UKST}
        {UKST}
\newcommand{\SDSS}
        {SDSS}

\newcommand{\samethanks}
	{{\Huge $^\star$}}

\newcommand{\etal}
	{et al.}
\newcommand{\eg}
	{e.g.}
\newcommand{\cf}
	{c.f.}
\newcommand{\ie}
	{i.e.}
\newcommand{\eq}[1]
	{equation~(\ref{equation:#1})}

\newcommand{\Eq}[1]
        {Equation~(\ref{equation:#1})}

\newcommand{\sect}[1]
	{Section~\ref{section:#1}}

\newcommand{\sects}[1]
        {Sections~\ref{section:#1}}

\newcommand{\fig}[1]
	{Fig.~\ref{figure:#1}}

\newcommand{\Fig}[1]
        {Fig.~\ref{figure:#1}}
\newcommand{\Figs}[1]
        {Figs.~\ref{figure:#1}}
\newlength{\singlefigureheight}
\newlength{\doublefigureheight}
\newlength{\triplefigureheight}
\newlength{\squarefigureheight}
\newcommand{\AaA}
        {A\&A}

\newcommand{\AJ}
        {AJ}
\newcommand{\ApJ}
        {ApJ}

\newcommand{\ARAA}
        {ARA\&A}

\newcommand{\MNRAS}
        {MNRAS}
\newcommand{\PhilTransA}
        {Phil.\ Trans.\ of the Royal Soc.\ A}

\newcommand{\Nature}
        {Nature}

\newcommand{\Science}
        {Science}

\newcommand{\omo}
        {\Omega_{\rmn m_0}}

\newcommand{\olo}
        {\Omega_{\Lambda_0}}

\newcommand{\FL}
        {FL}

\begin{document}

\title[Wide-separation lenses]
{The statistics of wide-separation lensed quasars}

\author[D.\ J.\ Mortlock and R.\ L.\ Webster]
       {
        Daniel J.\ Mortlock$^{1,2,3}$\thanks{
	E-mail: mortlock@ast.cam.ac.uk (DJM);
		rwebster@physics. unimelb.edu.au (RLW)}
	and Rachel L.\ Webster$^1$\samethanks\ \\
        $^1$School of Physics, The University of Melbourne, Parkville,
        Victoria 3052, Australia \\
	$^2$Astrophysics Group, Cavendish Laboratory, Madingley Road,
        Cambridge CB3 0HE, U.K. \\
	$^3$Institute of Astronomy, Madingley Road, Cambridge
	CB3 0HA, U.K. \\
       }

\date{
Accepted. 
Received; in original form 2000 April 19}

\pagerange{\pageref{firstpage}--\pageref{lastpage}}
\pubyear{2000}

\label{firstpage}

\maketitle

\begin{abstract}
The absence of any wide-separation gravitational lenses in the 
Large Bright Quasar Survey
is used to place limits on the population
of cluster-sized halos in the universe, and hence constrain
a number of cosmological parameters.
The results agree with previous investigations in
strongly ruling out the standard cold dark matter model but they
are consistent
with low-density universes in which the primordial fluctuation
spectrum matches both cluster abundances and
cosmic microwave background measurements.
These conclusions are essentially independent of the cosmological
constant, which is in stark contrast to the statistics of 
galaxy lenses.
The constraints presented here are nullified if clusters
have core radii of $\ga 10$ kpc,
but are free of a number of
potential systematic errors, due to the homogeneity of the data.
\end{abstract}


\section{Introduction}
\label{section:intro_lbqs}

The fraction of quasars which are gravitationally-lensed 
by intervening mass concentrations is a function of the deflector
population and the underlying cosmological model, which can
then be constrained from the observed lensing frequency
(\eg\ Kochanek 1993, 1996).
In most cases of quasar lensing known to date the 
principal deflectors are elliptical galaxies (\eg\ Keeton,
Kochanek \& Seljak 1997),
and the number of such lenses, together with the distribution of 
image separations and deflector redshifts,
has been used to place quite stringent limits on the galaxy population
(\eg\ Kochanek 1993) and the cosmological model (\eg\ 
Kochanek 1996; Falco, Kochanek \& Mu\~{n}oz 1998).
Unfortunately, the dependencies of the expected lensing probability 
on the galaxy population and the cosmology cannot 
be easily separated.
For instance, cosmological constant-dominated models
over-predict the number of lenses
if the local and cosmological galaxy populations are similar,
but can be made consistent with observations if 
high-redshift galaxies are optically thick due to dust 
(\eg\ Kochanek 1996; Malhotra, Rhoads \& Turner 1997).

One way to bypass the lack of knowledge about the high-redshift
galaxy population is to concentrate on lenses with larger
image separations (\ie\ $\ga 3$ arcsec). These
can only be produced by groups and clusters of galaxies, 
a population which is more closely linked to the underlying 
cosmological model.
Just four such wide-separation lenses have been confirmed:
Q~0957+561 (Walsh, Carswell \& Weymann 1979);
MG~2016+112 (Lawrence \etal\ 1984);
HE~1104$-$1805 (Wisotzki \etal\ 1993);
and
RXJ~0911+0551 (Bade \etal\ 1997).
There are more than ten other candidates,
but statistical arguments 
suggest that almost all of them are physically distinct binary quasars
(Kochanek, Falco \& Mu\~{n}oz 1999; Mortlock, Webster \&
Francis 1999).
If this interpretation is correct, less than 0.1 per cent 
of all quasars are lensed 
with such large image separations.
Conversely, theoretical calculations based on $N$-body simulations (\eg\
Wambsganss \etal\ 1995)
and analytical
models (\eg\ Narayan \& White 1988; Kochanek 1995;
Mortlock, Hewett \& Webster 1996)
predict more wide-separation lenses than are observed, 
if a standard cold dark matter (\cdm) cosmology is assumed.
However models with either a non-zero cosmological constant
or strong biasing (allowing the underlying dark matter distribution
to be much smoother than the galaxy distribution)
are consistent
with both lensing results and cosmic microwave background
(\cmb) anisotropies (Kochanek 1995). 

The main limitations on these results are the lack of knowledge 
of cluster core radii (If they are greater than $\sim 10$ kpc,
the standard models can be reconciled with the data.) and the 
heterogeneous nature of quasar catalogues used in previous analyses.
Both these points are addressed here, with 
the use of a homogeneous lens survey (\sect{lbqs_lensing_desc}),
and the possibility of a finite core included 
explicitly in the lens model (\sect{lensing_lbqs_calc}).
The absence of any lenses in the data can then be used to 
directly constrain a number of cosmological parameters
(\sect{results_lbqs_lens}). The accuracy of these inferences is
limited by both the simple model used for the deflector population
and the size of the quasar sample, 
as discussed in \sect{conc_lbqs_lens}.

\section{The Large Bright Quasar Survey}
\label{section:lbqs_lensing_desc}

The Large Bright Quasar Survey (LBQS; Hewett, Foltz \& Chaffee 1995)
is a sample of 1055 quasars taken from 18 
United Kingdom Schmidt Telescope (\UKST) fields.
The resultant magnitude-limited object catalogues were then used
in combination with objective-prism plates to generate 
a list of quasar candidates. As such, the quasar catalogue
was generated without any explicit morphological selection.
The magnitude limit\footnote{All magnitudes are in the 
$B_{\rmn J}$ system, but this subscript is omitted for brevity.}
varies from plate to plate,
ranging from $m_{\rmn lim} = 18.41$ to $m_{\rmn lim} = 18.85$,
and there is also a low redshift cut-off at $z = 0.2$.

The specific form of the differential number counts used in
the calculation of the lensing magnification bias (See \sect{p_lbqs}.) is
\begin{equation}
\label{equation:d2qdmdz}
\frac{{\rmn d}^2 N_{\rmn q}}{{\rmn d} m {\rmn d}z} \propto 
\frac{1}{10^{\alpha_{\rmn q}(m - m_{\rmn q0})}
     - 10^{\beta_{\rmn q}(m - m_{\rmn q0})}},
\end{equation}
where
$m_{\rmn{0}} = 19.0 \pm 0.2$ is the quasar break magnitude,
$\alpha_{\rmn q} = 0.9 \pm 0.1$ the bright-end slope
and 
$\beta_{\rmn q} = 0.3 \pm 0.1$ the faint-end slope.
This functional form is taken from 
Boyle, Shanks \& Peterson (1988) and Kochanek (1996),
but the parameter values are chosen to match the
\LBQS\ (for $m \la 19$) and the faint-end slope 
of the compilation of data presented by 
Hartwick \& Schade (1990).
Whilst Hewett, Foltz \& Chaffee (1993) found some discrepancies
between the \LBQS\ counts and the Boyle \etal\ (1988) sample, 
the simple parameterisation of \eq{d2qdmdz} is sufficiently
accurate in the context of this calculation.

As part of the \LBQS\ there was a systematic search for companions
within $\sim 10$ arcsec of each quasar.
The quality of the data used -- the same 
\UKST\ plates --
is such that most companions with $m_{B_{\rm J}} \la 21.5$ are found 
(Hewett \etal\ 1998),
but the search was incomplete for image separations
$\Delta \theta \la 3$ arcsec
due to the point spread function of the plates.
Given the \LBQS\ magnitude limit of $\sim 18.5$,
the companion search has sufficient
dynamic range to easily pick out most secondary lensed images,
and so can be considered complete in the annulus 
between 3 arcsec and 10 arcsec.
It is possible for galactic lenses to produce
image separations in this range, but only if they are both nearby and very
massive, and thus easily detectable.
Hence the \LBQS\ can be used to unambiguously 
constrain the population of galaxy group and cluster lenses.

The search for neighbouring images has yielded five quasar pairs
so far, and it is
`unlikely that further pairs will be identified' (Hewett \etal\ 1998).
Two of these pairs have vastly different redshifts, and are not the 
result of lensing (although their existence is evidence of the 
effectiveness of the companion search); the other three are either 
lenses or physical binary quasars.
Q~1009$-$0252 (Hewett \etal\ 1994) is a gravitational lens, but,
with $\Delta \theta = 1.5$ arcsec, it cannot be included 
in the lens calculation as the small-separation
lens search is not well characterised.
The second pair, Q~1429$-$0053 (Hewett \etal\ 1989) is a potential
wide-separation lens, with an image separation of $5.1$ arcsec
and apparently similar spectra.
However principal components analysis
(\eg\ Murtagh \& Hecht 1987) shows that the two spectra
are no more alike than a pair of spectra chosen at random from 
the survey (Mortlock \etal\ 1999). Combined with the absence of any potential
deflector and the statistical arguments of Kochanek \etal\ (1999),
it is highly unlikely that Q~1429$-$0053 is a lens.
The third pair, Q~2153$-$2056 (Hewett \etal\ 1998), has an image 
separation of $7.8$ arcsec, but very different spectra,
and is even less likely to be a lens. 
Thus the \LBQS\ almost certainly 
represents a sample of over 1000 quasars that is devoid of
wide-separation lenses.

\section{The calculation}
\label{section:lensing_lbqs_calc}

A given world model can be characterised by one
number:
$P_0$, the probability that no wide-separation lenses are
observed in the \LBQS. This, in turn, is simply the 
product of the probabilities that each 
individual quasar is unlensed, so that
\begin{equation}
\label{equation:P_0}
P_0 = \prod_{q = 1}^{N_{\rm q}} (1 - p_q) 
\simeq (1 - \langle p_{\rmn q} \rangle)^{N_{\rm q}},
\end{equation}
where $N_{\rm q}$ is the number of quasars in the survey,
$p_q$ is the probability that the $q$th quasar is lensed,
and $\langle p_{\rmn q} \rangle$ is the lensing probability averaged
over the survey.
Note that this is a cumulative probability, and so the
normalisation is unambiguous.

The likelihood of a given model is identified with $P_0$, thus ignoring
any prior information.
Hence, from \eq{P_0}, the only models that can be rejected at 
the 99 per cent level by the absence of lenses in the \LBQS\ are
those for which $\langle p_{\rmn q} \rangle \ga 0.004$.
This is true of a number of popular cosmological 
scenarios, and so the data at hand are far from redundant in 
this context.

Having set out the statistical framework for the calculation, 
the next step is to define the populations and 
mass distributions (\sects{model_lbqs_lensing} and 
\ref{section:deflectors_lbqs_lensing}, respectively),
from which the lensing probability can be calculated
(\sect{p_lbqs}).

\subsection{Deflector population}
\label{section:model_lbqs_lensing}

The population of collapsed halos can most accurately be estimated
from $N$-body simulations (\eg\ White \etal\ 1987; Efstathiou \etal\ 1988),
but this approach is too computationally expensive to explore a wide
range of cosmological models\footnote{The cosmological model is
defined by
$\omo$, the present day normalised density of the universe,
$\olo$, the current value of the similarly-normalised
cosmological constant,
and Hubble's constant,
$H_0 \simeq 70$ km s$^{-1}$ Mpc$^{-1}$.}.
The obvious alternative is 
the analytical Press-Schechter (1974) formalism 
(\eg\ Peacock 1999).
The halo population is normally given as a co-moving mass function, 
but the isothermal sphere lens model used in \sect{deflectors_lbqs_lensing} 
is parameterised by its line-of-sight velocity dispersion, $\sigma$,
rather than mass.
Equating the total mass to that inside the virial radius of the 
isothermal sphere (\eg\ Narayan \& White 1988; Kochanek 1995) 
results in the conversion
\begin{equation}
\label{equation:sigma_m}
M(\sigma, z) = \frac{\left( 2^{1/2} \sigma \right)^3}
{10 G H(z)},
\end{equation}
where $G$ is Newton's constant and 
$H(z) = H_0
[
\omo (1 + z)^3 + \olo - (\omo + \olo - 1) (1 + z)^2
]^{1/2}$.
The final expression for the co-moving halo population is then
(Mortlock 1999)
\begin{eqnarray}
\label{equation:dndsig_ps}
\frac{{\rm d} n_{\rm d}}{{\rm d} \sigma}(z)
& = & - \frac{15 \Omega_{\rm m_0}H_0^3}{8 \pi^{3/2}}
\frac{H(z) / H_0 \, \delta_{\rm crit}(z)}
{\sigma^4\, \Delta \left[
M(\sigma, z)
\right]} \nonumber \\ 
& \times &
\frac{{\rm d} \ln\left\{\Delta\left[
M(\sigma, z)
\right]
\right\}}{{\rm d} \ln(\sigma)} 
\exp\left\{ \frac{\delta^2_{\rm crit}(z)}
{2 \Delta^2 \left[
M(\sigma, z)
\right]} \right\}
, 
\end{eqnarray}
where $M(\sigma, z)$ is given in \eq{sigma_m}.
The cosmological model enters \eq{dndsig_ps}
only through
$\delta_{\rmn crit}(z)$, the extrapolated linear overdensity
that would have collapsed at redshift $z$ (\eg\ Peebles 1980).
The present day variance on mass scale $M$, $\Delta(M)$,
is determined by the power spectrum of density fluctuations.
The approximate \cdm\ power spectrum of 
Efstathiou, Bond \& White (1992) is adopted here. 
Most of the power spectrum parameters have little 
influence on the lensing likelihood
(Kochanek 1995; Mortlock 1999), 
and so only the primordial power-law slope, $n$, and
the normalisation, $\Delta_8$ 
(the present day variance in spheres of radius 8 Mpc),
are allowed to vary.

\subsection{Lens model}
\label{section:deflectors_lbqs_lensing}

The mass distributions of galaxy clusters -- the inner regions
in particular -- are not well constrained by observations or theory:
\cdm-based $N$-body simulations suggest that they are singular
(\eg\ Navarro, Frenk \& White 1997);
the properties
of giant arcs imply they have small core radii
(\eg\ Fort \etal\ 1992; Smail \etal\ 1995,
but see also Bartelmann 1996) or very massive central
galaxies (Williams, Navarro \& Bartelmann 1999);
and other observational data span
the possibilities (\eg\ Narayan, Blandford \& Nityananda 1984;
Mohr \etal\ 1996; Pointecouteau \etal\ 1999).
A simple isothermal sphere
(\eg\ Turner, Ostriker \& Gott 1984; Binney \& Tremaine 1987)
is adopted here because
the principal analytic alternative -- the Navarro \etal\ (1997)
profile -- is inconsistent with the 
strong lensing properties of clusters (Williams \etal\ 1999).
However the main short-coming of both these 
models is the absence of any substructure, the inclusion of 
which increases image separations by $\sim 50$ per cent 
on average (Bartelmann, Steinmetz \& Weiss 1995).

The isothermal profile, parameterised primarily
its velocity dispersion, is assumed to have a core radius,
$r_{\rmn c}$, which scales as
$r_{\rm c} = r_{\rm c*} (\sigma/\sigma_*)^{u_{\rm c}}$,
where 
$\sigma_* = $ 1000 km s$^{-1}$ is chosen arbitrarily.
Its lensing properties are usually cast in terms of the singular 
model's Einstein radius, $\theta_{\rmn E}$ 
(which is dependent on the cosmological model; 
Schneider, Ehlers \& Falco 1992), 
and the critical radius in the lens
plane, $\beta_{\rmn crit}$ (which decreases as the core radius increases).
For the definitions of these terms and a further discussion of this 
lens model see, 
Hinshaw \& Krauss (1984), Kochanek (1996) or 
Mortlock \& Webster (2000).

\subsection{Lensing probability}
\label{section:p_lbqs}

From the properties of the lens model it is possible to calculate
the probability, $p_{\rmn q}$, that a quasar of magnitude $m_{\rmn q}$ 
(in the parent survey)
and redshift $z_{\rmn q}$ is found to be lensed. This is derived 
rigorously for a typical lens survey in 
Kochanek (1996) and Mortlock \& Webster (2000),
but there are some important differences in the calculation
of the magnification bias here.
In most lens surveys, the angular separation of the images 
is less than the angular resolution of the parent survey, 
and so, for a lensed quasar, $m_{\rmn q}$ is given by summing
the fluxes of all the images. 
However, in the case of the \UKST\ data from which the \LBQS\ was selected,
only images separated by $\la 5$ arcsec 
appear merged (Webster, Hewett \& Irwin 1988, although
Hewett \etal\ 1995 give $\sim 6$ arcsec; this discrepancy leads to  
a $\sim 10$ per cent uncertainty in the calculated lensing probability), 
whereas the 
search annulus of the lens survey extends from $\Delta \theta_{\rmn min}
= 3$ arcsec to $\Delta \theta_{\rmn max} = 10$ arcsec.
This is an unusual situation -- the lens survey improves on the
parent survey in depth, not resolution.
For any lenses with $\Delta \theta \la 5$ arcsec the standard
magnification bias is correct, but for wider-separation multiples
$m_{\rmn q}$ is determined only by the magnification of the 
brightest individual image\footnote{Merging images are 
not explicitly treated here.}.
This is often considerably less than the total
magnification, and the resultant reduction in the lensing probability can 
be quite marked, as shown below. 

The first step 
in evaluating $p_{\rmn q}$ is to calculate $p_{\rmn q,d}$, 
the probability that a given quasar is lensed by a particular
halo, defined by its redshift, $z_{\rmn d}$ and its velocity dispersion,
$\sigma$. With the normal total magnification bias, this is given by
\begin{equation}
\label{equation:p_gq_normal}
p_{\rmn q,d} =
\frac{
\int_0^{\beta_{\rmn crit}}
2 \pi \beta \,S(\beta)
\left. \frac{{\rmn d}^2N_{\rmn q}}{{\rmn d} z_{\rmn q}\, {\rmn d} m}
\right|_{m = m_{\rmn q} + 5/2\, \log[ \mu_{\rmn tot} (\beta)]}
\, {\rmn d} \beta}
{4 \pi \, \frac{{\rmn d}^2N_{\rmn q}}
{{\rmn d} z_{\rmn q}\, {\rmn d} m_{\rmn q}}},
\end{equation}
where
the quasar luminosity function, ${\rmn d}^2N_{\rmn q}/
{\rmn d} z\, {\rmn d} m$, is given in \eq{d2qdmdz} and
$S(\beta)$ is the selection function (\cf\ Kochanek 1996).
The latter is approximated by
\begin{eqnarray}
S(\beta) & = & H \! \left[\Delta m_{\rmn max} - \Delta m(\beta)\right]
\nonumber \\
& \times &
H \! \left[\Delta \theta (\beta) - \Delta \theta_{\rmn min} \right]\,
H \! \left[\Delta \theta_{\rmn max} - \Delta \theta (\beta) \right],
\end{eqnarray}
where $H(x)$ is the Heavyside step function.

\begin{figure}
\includegraphics{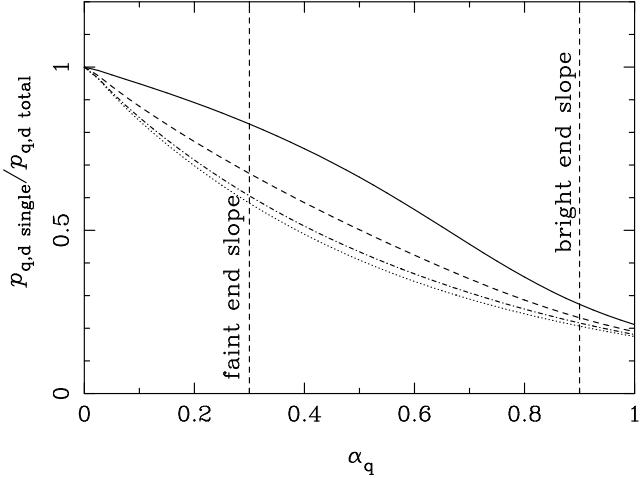}
\vspace{\singlefigureheight}
\caption{The effect of the single image magnification bias
(relative to the standard total magnification bias) on the
lensing probability of a single deflector, $p_{\rmn q,d}$,
as a function of the logarithmic slope of the source
counts, $\alpha_{\rmn q}$.
The bright end and faint end slopes inferred from the
Boyle \etal\ (1988) quasar sample are also shown.
The deflector (which has a velocity dispersion of 
$\sigma = $ 1000 km s$^{-1}$)
and source are at redshifts of 0.5 and 2, respectively,
and $\omo = 1$ and $\olo = 0$ is assumed.
Results are shown for several core radii:
$r_{\rmn c} = 0$ kpc (singular model; solid line);
$r_{\rmn c} = 5$ kpc (dashed line);
$r_{\rmn c} = 10$ kpc (dot-dashed line)
and
$r_{\rmn c} = 20$ kpc (dotted line).}
\label{figure:p_lens_mb}
\end{figure}

\Eq{p_gq_normal} can be converted to the 
single image magnification
bias simply by replacing $\mu_{\rmn tot}(\beta)$
with the magnification of the brightest
image, $\mu_{\rmn max}(\beta)$.
As $\mu_{\rmn max} < \mu_{\rmn tot}$ in all cases of multiple
imaging, the single image magnification bias always 
results in a lower value of $p_{\rmn q,d}$.
The ratio of the two probabilities is particularly simple to 
calculate
if the deflectors are assumed to be singular: the 
integration variable in \eq{p_gq_normal} can be changed
to 
$\mu_{\rmn tot} = 2 / (\beta/\theta_{\rmn E})$,
and
$\mu_{\rmn max} = (1 + \beta/\theta_{\rmn E})
/ (\beta/\theta_{\rmn E})$.
Thus
\[
\frac{p_{\rmn q,d\, single}}{p_{\rmn q,d\, total}} =
\frac{
\int_2^\infty
\frac{2}{(\mu_{\rmn max} - 1)^3}
\left. \frac{{\rmn d}^2N_{\rmn q}}{{\rmn d} z_{\rmn q}\, {\rmn d} m}
\right|_{m = m_{\rmn q} + 5/2\, \log( \mu_{\rmn max})}
\!\! {\rmn d} \mu_{\rmn max}
}
{
\int_2^\infty
\frac{8}{\mu_{\rmn tot}^3}
\left. \frac{{\rmn d}^2N_{\rmn q}}{{\rmn d} z_{\rmn q}\, {\rmn d} m}
\right|_{m = m_{\rmn q} + 5/2\, \log( \mu_{\rmn tot})}
\!\! {\rmn d} \mu_{\rmn tot}
},
\]
\begin{equation}
\label{equation:p_lens_mb}
\end{equation}
where $\Delta m_{\rmn max} \rightarrow \infty$ for simplicity.
\Fig{p_lens_mb} shows how this ratio varies with the slope of 
the quasar number counts, $\alpha_{\rmn q}$,
if they are given by a single 
power law, ${\rmn d}N_{\rmn q}/{\rmn d} m \propto 10^{\alpha_{\rmn q} m}$.
The main difference between the two situations is the reduction in 
the amplitude of the high-magnification tail in the case of 
the single image magnification bias, which in turn has a greater
effect on $p_{\rmn q,d}$ if the luminosity function is steeper.
Thus the wide-separation lensing probability for bright
quasars (including the \LBQS) is up to a factor of three lower than previous 
calculations would suggest\footnote{Most relevantly,
Kochanek (1995) used the normal
magnification bias,
but due to the image separations under consideration (up to 1 arcmin),
the weaker, single image bias would have given a more
realistic value of the lensing probability.
For example, the two standard deviation
result that $0.27 \la \Delta_8 \la 0.63$ (in the
standard \cdm\ model) becomes
$0.35 \la \Delta_8 \la 0.75$.}.

The probability that a quasar is lensed by any halo is obtained by
integrating $p_{\rmn q,d}$ over the deflector population, to give
\begin{equation}
\label{equation:p_q_lbqs}
p_{\rmn q} = \int_0^{z_{\rmn q}} \int_0^\infty
\left. \frac{{\rmn d}V_0}{{\rmn d} z}\right|_{z = z_{\rmn d}}
\frac{{\rmn d}n_{\rmn d}}{{\rmn d}\sigma} (z) \,
p_{\rmn q,d} 
\, {\rmn d}\sigma \, {\rmn d}z_{\rmn d},
\end{equation}
where ${\rmn d}n_{\rmn d}/{\rmn d}\sigma(z)$, 
the co-moving density of deflectors, 
is given in \sect{model_lbqs_lensing}.

\begin{figure}
\includegraphics{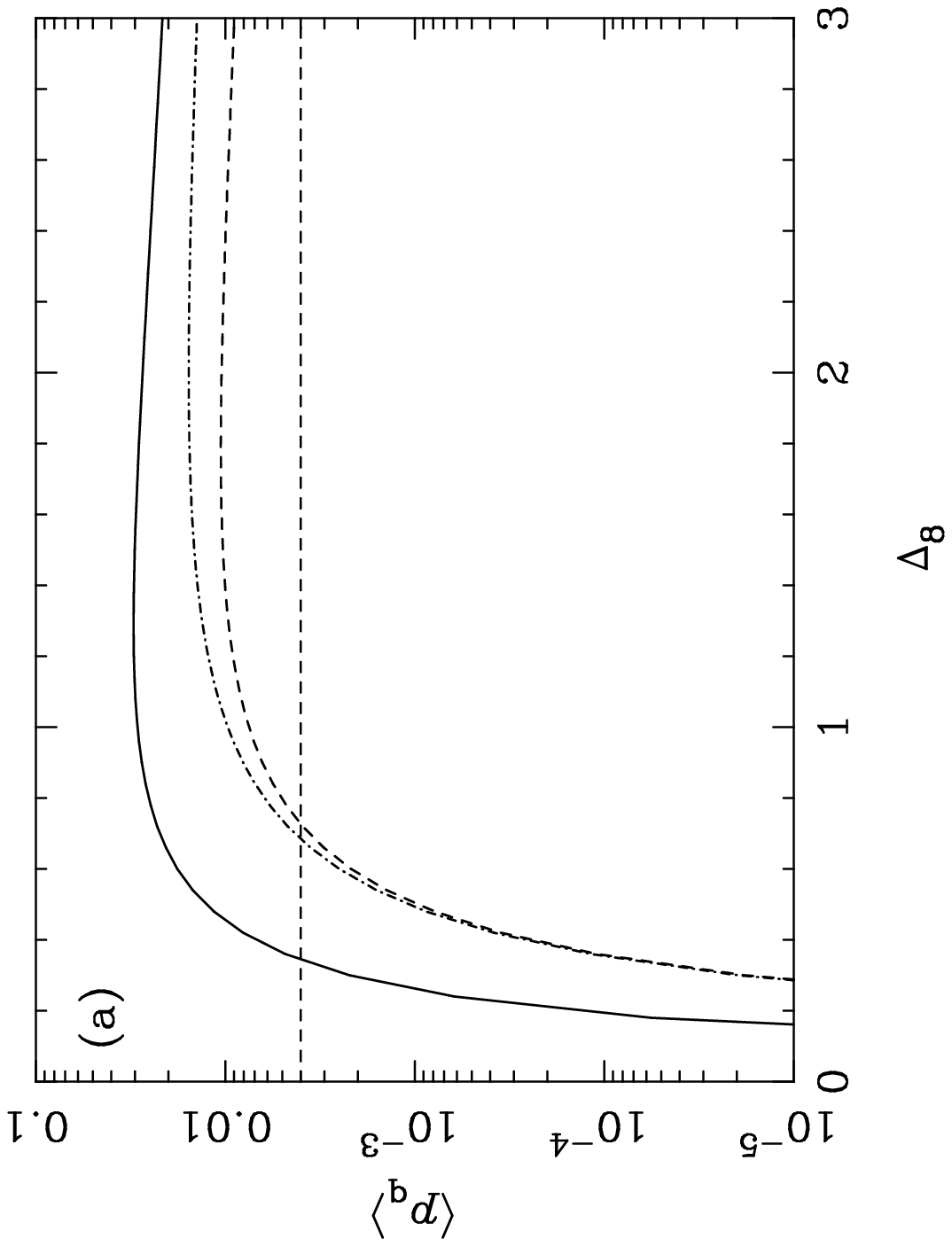}
\includegraphics{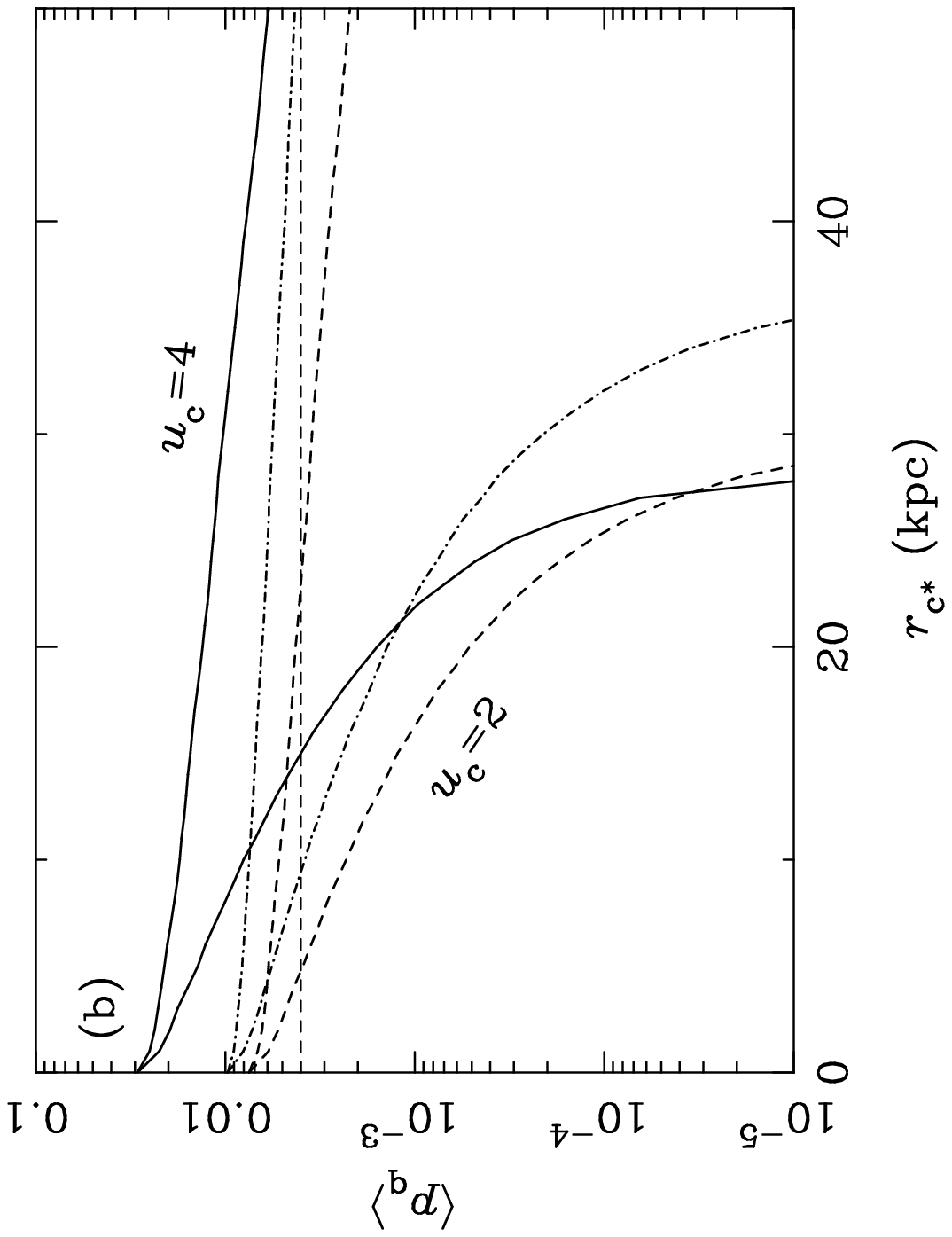}
\includegraphics{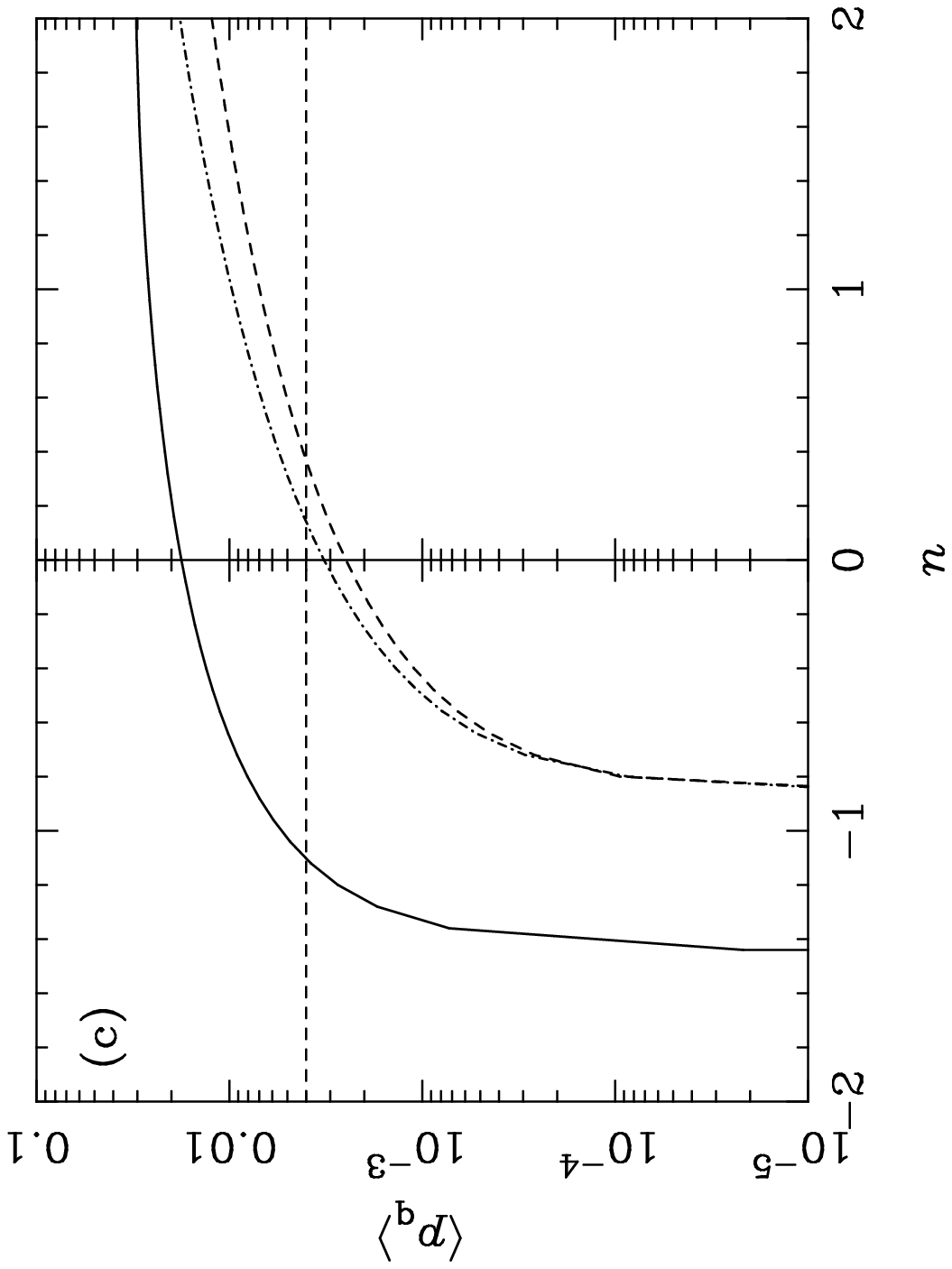}
\vspace{\triplefigureheight}
\caption{The average probability
that any quasar in the \LBQS\ is observed
to be lensed, $\langle p_{\rmn q} \rangle$.
The dependence on the normalisation, $\Delta_8$, is shown in (a);
the variation with the canonical core radius, $r_{\rmn c*}$
is shown in (b); and
the dependence on the slope of the power spectrum, $n$,
is shown in (c).
The `default' values are $\Delta_8 = 1$, $r_{\rmn c*} = 0$ and
$n = 1$
and the three lines represent different
cosmological models:
$\Omega_{\rm m_0} = 1.0$ and $\Omega_{\Lambda_0} = 0.0$ (solid lines);
$\Omega_{\rm m_0} = 0.3$ and $\Omega_{\Lambda_0} = 0.0$ (dashed lines);
and
$\Omega_{\rm m_0} = 0.3$ and $\Omega_{\Lambda_0} = 0.7$ (dot-dashed lines).
Models with $\langle p_{\rmn q} \rangle > 0.004$
(the horizontal dashed line) can be rejected at the 
99 per cent level.}
\label{figure:p_lens_ps_lbqs}
\end{figure}

\begin{figure}
\includegraphics{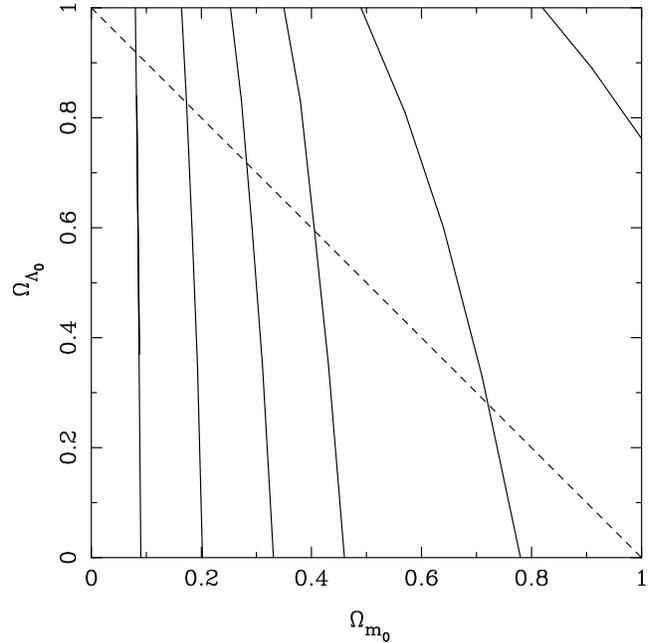}
\vspace{\squarefigureheight}
\caption{The probability that any \LBQS\ quasar is lensed,
$\langle p_{\rmn q} \rangle$, as a function of
$\Omega_{\rm m_0}$ and $\Omega_{\Lambda_0}$.
The power spectrum is parameterised by $\Delta_8 = 1$,
and $n = 1$, and the lenses are singular.
The contours start at $\langle p_{\rmn q} \rangle = 0$
(the $\Omega_{\rm m_0} = 0$ axis)
and then increase in steps of 0.005; the
highest contour (at the top-right) is
$\langle p_{\rmn q} \rangle = 0.03$.
The dashed line denotes spatially flat models.}
\label{figure:om_lam_ps}
\end{figure}

This must be calculated separately for each quasar in the 
\LBQS, due mainly to their large redshift range, and
the resultant likelihoods can be averaged to
give $\langle p_{\rmn q} \rangle$.
This is shown in \Figs{p_lens_ps_lbqs} and \ref{figure:om_lam_ps}
as function of several model parameters.
In \fig{p_lens_ps_lbqs} (a),
the lensing probability initially increases very rapidly with $\Delta_8$,
as the velocity dispersion of the largest halos 
depends exponentially on $\Delta_8$ (Kochanek 1995).
For $\Delta_8 \ga \delta_{\rmn crit}$, however, 
the lensing likelihood flattens off as
the largest halos produce image separations of $\ga$ 10 arcsec,
and so cannot contribute to $\langle p_{\rmn q} \rangle$.
As shown in \fig{p_lens_ps_lbqs} (b),
the number of lenses decreases with cluster 
core radii, as expected (\eg\ Kochanek 1995).
However the dependence is very weak if $u_{\rmn c} \simeq 4$.
For a given $r_{{\rmn c}*}$, the smaller deflectors (which are capable of 
producing lenses with $\Delta \theta \simeq$ 3 arcsec) are 
more effective, being nearly singular. Unfortunately this 
degeneracy
means that any limits placed on the other model parameters 
are weakened by the fact that $u_{\rmn c}$ is essentially 
unconstrained. 
The situation with respect to $n$ is more promising -- 
as can be seen from \fig{p_lens_ps_lbqs} (c) the 
slope of the power spectrum has only a small effect on the 
lens statistics.

The dependence of $\langle p_{\rmn q} \rangle$ on the cosmological
model is illustrated in both \fig{p_lens_ps_lbqs} and \fig{om_lam_ps}.
The most striking aspect of these plots is that 
$\omo$ is considerably more important than $\olo$,
whereas it is the cosmological constant that dominates the
statistics of lensing by galaxy-scale objects.
For a given $\omo$
the volume element is greater in high-$\olo$ universes,
but clusters form at later times, so the increase of 
$\langle p_{\rmn q} \rangle$ with $\olo$ is only mild.
The strong dependence on $\omo$ comes about primarily
as halos are heavier for a fixed $\Delta_8$.
A cluster that has collapsed from 
an initial perturbation of a given co-moving scale
has mass $M \propto \omo$, velocity dispersion $\sigma \propto \omo^{1/3}$
[from \eq{sigma_m}], and hence a lensing cross-section that increases
as $\omo^{4/3}$ (\eg\ Turner \etal\ 1984; Kochanek 1995).
This dependence is apparent for low-density models (See \fig{om_lam_ps}.),
but as $\omo \rightarrow 1$ the predominant halos have typical 
image separations that are greater than $\Delta \theta_{\rmn max}$.
A lens survey with a broader search annulus than that of the \LBQS\
would be required to probe models with either $\Delta_8$ 
or $\omo$ of order unity.

\section{Results}
\label{section:results_lbqs_lens}

\begin{figure}
\includegraphics{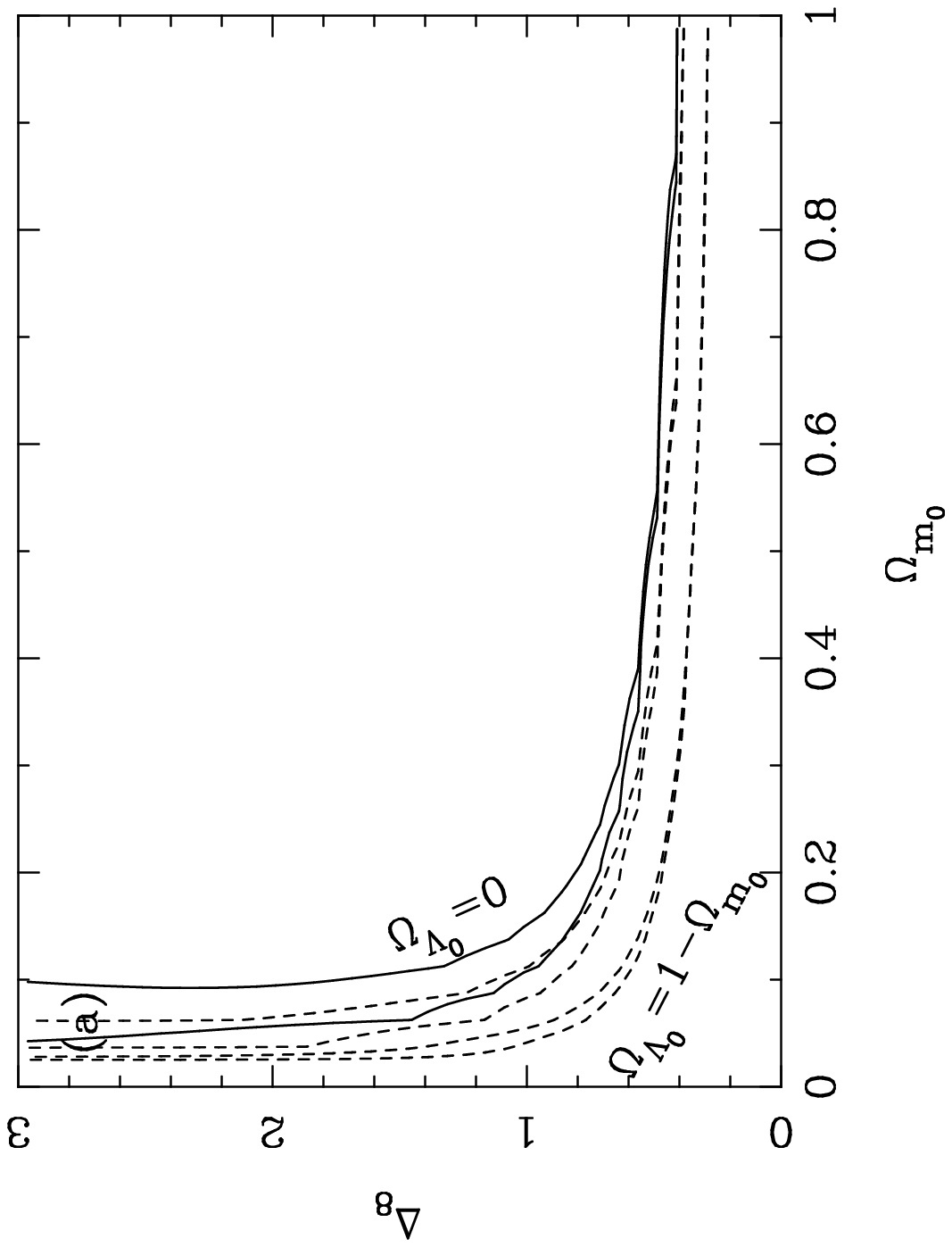}
\includegraphics{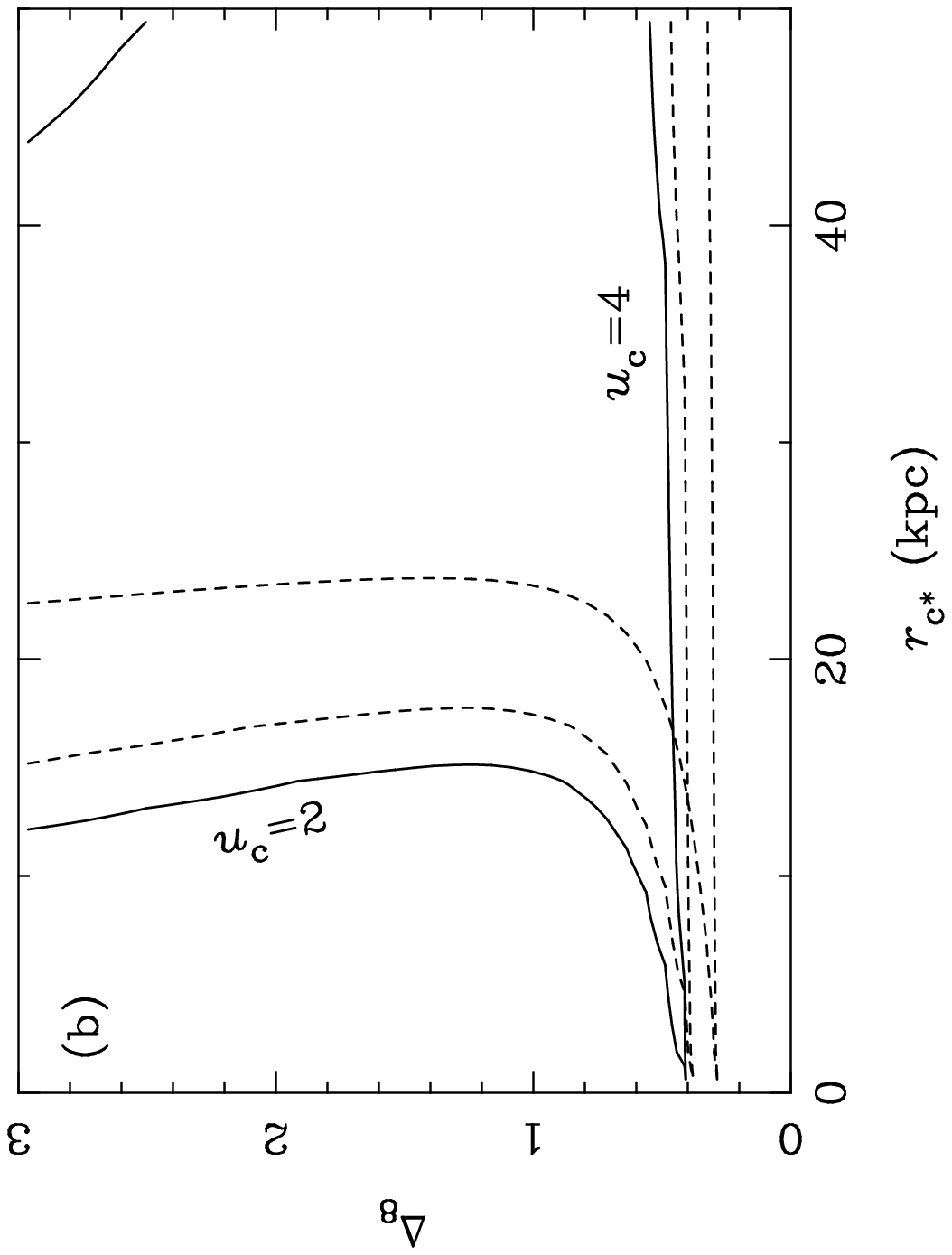}
\includegraphics{n_var8.ps}
\vspace{\triplefigureheight}
\caption{The constraints placed on 
the cosmological model and the power spectrum parameters from
the absence of lenses in the \LBQS. 
The solid contours are at
$P_0 = 0.01$; the dashed contours are at
$P_0 = 0.05$ and $P_0 = 0.5$.
In all panels the vertical axis gives the normalisation 
of the power spectrum, $\Delta_8$. 
The horizontal axes are: 
$\Omega_{\rm m_0}$ (for both $\Omega_{\Lambda_0} = 0$ and flat models) in (a);
the canonical core radius, $r_{\rmn c*}$ (for $u_{\rmn c} = 2$
and $u_{\rmn c} = 4$) in (b); and
the slope of the power spectrum, $n$, in (c).
In panels (b) and (c) an \EdS\ cosmology is assumed,
and $r_{\rmn c*} = 0$ and
$n = 1$ are assumed except where
these parameters are explicitly varied.}
\label{figure:p_model_ps}
\end{figure}

The selection criteria for a lens to appear in the \LBQS\
(\sect{lbqs_lensing_desc}) and the lensing calculation
described in \sect{lensing_lbqs_calc} can now be combined to 
give $P_0$, the probability that the \LBQS\ contains no 
wide-separation lenses. 
This is shown for several combinations of model parameters 
in \fig{p_model_ps}. The contours shown are for 
$P_0 = 0.01$ (\ie\ 99 per cent limits),
$P_0 = 0.05$ and $P_0 = 0.5$, but are only one-sided as 
models which predict an arbitrarily low number of lenses
are perfectly consistent with the data.

\Fig{p_model_ps} (a) shows that $\Delta_8 \la 0.4$ 
(with 99 per cent confidence)
in an Einstein-de Sitter (\EdS) cosmology with a standard \cdm\ spectrum.
This is considerably lower than the value of $1.4 \pm 0.1$ 
inferred from the {\em Cosmic Background Explorer} 
(\COBE) data (Smoot \etal\ 1992).
Standard \COBE-normalised \cdm\ models can
be made consistent with the data if $r_{{\rmn c}*}$
$\ga 10$ kpc and $u_{\rmn c} \simeq 2$,
as shown in \fig{p_model_ps} (b).
Although such models are somewhat contrived\footnote{There must also be some
nearly-singular clusters to account for the morphology of the
confirmed wide-separation
lenses listed in \sect{intro_lbqs}.
Observations of giant arcs (\eg\ Smail \etal\ 1995; Bartelmann 1996)
also argue against such high values of $_{{\rmn c}*}$.},
the scaling of cluster core radii is so poorly constrained that
they cannot be completely ruled out.
However it is the same low-density models implied by 
observations of high-redshift supernov\ae\ 
(\eg\ Schmidt \etal\ 1998; Perlmutter \etal\ 1999) 
and the position of the first \cmb\ Doppler peak 
(\eg\ Efstathiou \etal\ 1999; de Bernardis \etal\ 2000)
that are favoured by the data at hand.
There may be a slight discrepancy if 
measurements of $\Delta_8$ are also considered, 
as the values inferred from 
the local cluster population
(\eg\ Peebles 1989; Frenk \etal\ 1990; Bahcall, Fan \& Cen 1997)
and the galaxy-galaxy correlation function
(\eg\ Maddox \etal\ 1990; Efstathiou \etal\ 1992) 
are slightly higher than allowed in the $\omo = 0.3$ model here.
However, if $\omo \simeq 0.2$ and the universe is flat then
the greatest source of disagreement with the absence of 
wide-separation lenses
in the \LBQS\ is, somewhat ironically, the low number of
small-separation lenses that preclude $\olo \simeq 0.8$ 
(\eg\ Kochanek 1996).

It is possible, although very unlikely, that there is a wide-separation 
lens in the \LBQS, as Q~1429$-$0053 has not
been conclusively proved to be a binary quasar and there is still a small
amount of the companion search to be completed (Hewett \etal\ 1998).
If such a lens were to be discovered then
the limits on the model parameters would be shifted
somewhat. 
Performing a calculation similar to that
described in \sect{lensing_lbqs_calc} gives the
probability of finding one lens in the \LBQS\ as
\begin{eqnarray}
P_1 & \simeq & N_{\rmn q} \langle p_{\rmn q} \rangle
(1 - \langle p_{\rmn q} \rangle)^{N_{\rmn q} - 1} \nonumber \\
& \simeq & N_{\rmn q} P_0^{(N_{\rmn q} - 1)/N_{\rmn q}}
(1 - P_0)^{1 / N_{\rmn q}} ,
\end{eqnarray}
where $N_{\rmn q} = 1055$ is the number of quasars in the survey.
From this conversion, the 1 per cent, 5 per cent
and 50 per cent contours for $P_0$
shown in \fig{p_model_ps}
become 5 per cent, 15 per cent
and 35 per cent contours for $P_1$, respectively.
In other words, the 99 per cent limits would be weakened 
to 95 per cent limits.
The constraints would also become two-sided, 
but the models that would be ruled out by the presence of a lens 
have $\langle p_{\rmn q} \rangle \simeq 10^{-5}$,
and would not be considered plausible a priori.
Thus the presence of a wide-separation lens in the \LBQS\ would 
broaden the constraints on the model parameters, but would
not qualitatively change the conclusions.

\section{Conclusions}
\label{section:conc_lbqs_lens}

The fact that no wide-separation ($\Delta \theta \ga 3$ arcsec)
lenses have been discovered in the \LBQS\ places tight
constraints on the population of cluster-mass objects in the universe.
Assuming a hierarchical theory of halo
formation, one-sided limits can then be placed on 
several cosmological parameters. 
Interestingly, the results presented in 
\sect{results_lbqs_lens} (See also Kochanek 1995.) indicate 
that only three model parameters are strongly constrained
($\omo$, $\Delta_8$ and $r_{{\rmn c}*}$), and that the lens
statistics are almost independent of the cosmological constant.
This is in stark contrast with the probability of lensing 
by galaxies, which is dominated by $\olo$.

The most straightforward result is that standard the \cdm\ model
(\ie\ an \EdS\ cosmology with a power spectrum index of 
$n = 1$) is completely ruled out unless the 
normalisation is very low ($\Delta_8 \la 0.4$ at the 
99 per cent level).
Only if the core radii of $\sigma \simeq$ 1000 km s$^{-1}$ clusters are 
$\ga 10$ kpc,
and $r_{\rmn c} \propto \sigma^2$
can these limits be avoided. 
For non-\EdS\ cosmologies, the lensing results are in agreement 
with most a priori reasonable models 
(as discussed in \sect{results_lbqs_lens}), although
$\Omega_{\rm m_0} \la 0.2$ is implied unless $\Delta_8$ is low 
or, again, clusters have large core radii.

Whilst the conclusions are in agreement with most expectations,
it is important to consider the various approximations in this calculation,
and their effects on the parameter constraints.
Fortunately, the lensing probability varies as $\Omega_{\rm m_0}^{4/3}$
and increases exponentially with $\Delta_8$ (provided $\Delta_8 \la 1$), 
whereas 
the use of the Press-Schechter (1974) formalism, 
the calculation of $\delta_{\rmn crit}$,
the choice of a smooth, spherical lens model,
and the treatment of the magnification bias
all lead to uncertainties of tens of per cent at most.
Hence the limits on
$\Omega_{\rm m_0}$ and $\Delta_8$ could be changed by similar amounts.
The choice of mass profile used for the
lenses is more important (as illustrated by the effect of 
$r_{{\rmn c}*}$) and it would be especially interesting to perform the
above calculation using the Navarro \etal\ (1997)
halo model, as the mass distribution of the halos in their 
prescription is assumed to be directly related to the
underlying cosmological model.

The \LBQS\ represents an ideal type of data-set for this 
kind of analysis,
as it has well-characterised
selection effects 
and 
the companion search
was both deep and systematic;
the only real shortcoming is the size of the survey.
Both the Jodrell Bank-Very Large Array Astrometric Survey
(\JVAS; Patnaik \etal\ 1992) of $\sim 2500$
sources and 
the smaller {\em Hubble Space Telescope} Snapshot Survey (Maoz \etal\ 1997) 
of $\sim 500$ quasars also contain no confirmed lenses with
$\Delta \theta \ga 5$ arcsec (Marlow \etal\ 1998), 
and so would imply similar one-sided limits to those presented here.
The situation should be improved with the advent of much larger
surveys, such as
the 2 degree Field (\TdF) quasar survey (\eg\ Boyle \etal\ 1999a,b),
with $\sim 3 \times 10^4$ quasars,
and the Sloan Digital Sky Survey
(\SDSS; \eg\ Szalay 1998; Loveday \& Pier 1998), 
with $\sim 10^5$ quasars.
The \TdF\ survey's companion search is not particularly deep, 
and will only include lenses with image separations 
greater than 8 arcsec. Nonetheless, the sheer size of 
the data-set will allow a lensing determination of 
$\Omega_{\rm m_0}$ and $\Delta_8$ to within
$\sim 10$ per cent.
The \SDSS\ quasar survey includes high-resolution imaging, 
and should thus provide a large sample of both galactic 
and wide-separation lenses.
This will allow $\Omega_{\rm m_0}$, $\Omega_{\Lambda_0}$ 
and $\Delta_8$ to be constrained simultaneously to within several per cent
from lensing statistics alone.

\section*{Acknowledgments}

Many thanks to both Paul Hewett, 
whose intimate understanding of the \LBQS\ was crucial to this analysis, 
and to the anonymous referee, 
whose comments improved the form of this paper.
DJM was supported by an Australian Postgraduate Award.

\bsp
\label{lastpage}
\end{document}